\documentstyle[12pt,epsf]{article}

\def \roots {\sqrt {s}}
\def \Pt {{P}_{T}}
\def \Et {{E}_{T}}
\def \lq {{\cal S}_1}
\def \mlq {M_{{\cal S}_1}}
\def \mej {M_{ej}}

\newcommand{\ppbar}{p\bar{p}}

\newcommand{\ttbar}{t\bar{t}}
\newcommand{\bbar}{b\bar{b}}
\newcommand{\lqlqbar}{{\cal S}_1 \bar{{\cal S}_1}}

\begin{document}

\begin{center}

{\bf Search for  First Generation Leptoquark Pair Production in $\ppbar$
Collisions at $\roots$ = 1.8 TeV.}

\font\eightit=cmti8
\def\r#1{\ignorespaces $^{#1}$}
\hfilneg
\begin{sloppypar}
\noindent
F.~Abe,\r {17} H.~Akimoto,\r {38}
A.~Akopian,\r {31} M.~G.~Albrow,\r 7 A.~Amadon,\r 5 S.~R.~Amendolia,\r {27} 
D.~Amidei,\r {20} J.~Antos,\r {33} S.~Aota,\r {36}
G.~Apollinari,\r {31}
% ...
\end{sloppypar}
\vskip .026in
\begin{center}
(CDF Collaboration)
\end{center}

\vskip .026in
\begin{center}
\r 1  {\eightit Argonne National Laboratory, Argonne, Illinois 60439} \\
\r 2  {\eightit Istituto Nazionale di Fisica Nucleare, University of Bologna,
I-40127 Bologna, Italy} \\
\r 3  {\eightit Brandeis University, Waltham, Massachusetts 02254} \\
\r 4  {\eightit University of California at Los Angeles, Los 
Angeles, California  90024} \\  
\r 5  {\eightit University of Chicago, Chicago, Illinois 60637} \\
\r 6  {\eightit Duke University, Durham, North Carolina  27708} \\
\r 7  {\eightit Fermi National Accelerator Laboratory, Batavia, Illinois 
60510} \\
\r 8  {\eightit University of Florida, Gainesville, FL  32611} \\
\r 9  {\eightit Laboratori Nazionali di Frascati, Istituto Nazionale di Fisica
               Nucleare, I-00044 Frascati, Italy} \\
\r {10} {\eightit University of Geneva, CH-1211 Geneva 4, Switzerland} \\
\r {11} {\eightit Harvard University, Cambridge, Massachusetts 02138} \\
\r {12} {\eightit Hiroshima University, Higashi-Hiroshima 724, Japan} \\
\r {13} {\eightit University of Illinois, Urbana, Illinois 61801} \\
\r {14} {\eightit Institute of Particle Physics, McGill University, Montreal 
H3A 2T8, and University of Toronto,\\ Toronto M5S 1A7, Canada} \\
\r {15} {\eightit The Johns Hopkins University, Baltimore, Maryland 21218} \\
\r {16} {\eightit Institut f\"{u}r Experimentelle Kernphysik, 
Universit\"{a}t Karlsruhe, 76128 Karlsruhe, Germany} \\
\r {17} {\eightit National Laboratory for High Energy Physics (KEK), Tsukuba, 
Ibaraki 315, Japan} \\
\r {18} {\eightit Ernest Orlando Lawrence Berkeley National Laboratory, 
Berkeley, California 94720} \\
\r {19} {\eightit Massachusetts Institute of Technology, Cambridge,
Massachusetts  02139} \\   
\r {20} {\eightit University of Michigan, Ann Arbor, Michigan 48109} \\
\r {21} {\eightit Michigan State University, East Lansing, Michigan  48824} \\
\r {22} {\eightit University of New Mexico, Albuquerque, New Mexico 87131} \\
\r {23} {\eightit The Ohio State University, Columbus, OH 43210} \\
\r {24} {\eightit Osaka City University, Osaka 588, Japan} \\
\r {25} {\eightit Universita di Padova, Istituto Nazionale di Fisica 
          Nucleare, Sezione di Padova, I-36132 Padova, Italy} \\
\r {26} {\eightit University of Pennsylvania, Philadelphia, 
        Pennsylvania 19104} \\   
\r {27} {\eightit Istituto Nazionale di Fisica Nucleare, University and Scuola
               Normale Superiore of Pisa, I-56100 Pisa, Italy} \\
\r {28} {\eightit University of Pittsburgh, Pittsburgh, Pennsylvania 15260} \\
\r {29} {\eightit Purdue University, West Lafayette, Indiana 47907} \\
\r {30} {\eightit University of Rochester, Rochester, New York 14627} \\
\r {31} {\eightit Rockefeller University, New York, New York 10021} \\
\r {32} {\eightit Rutgers University, Piscataway, New Jersey 08855} \\
\r {33} {\eightit Academia Sinica, Taipei, Taiwan 11530, Republic of China} \\
\r {34} {\eightit Texas A\&M University, College Station, Texas 77843} \\
\r {35} {\eightit Texas Tech University, Lubbock, Texas 79409} \\
\r {36} {\eightit University of Tsukuba, Tsukuba, Ibaraki 315, Japan} \\
\r {37} {\eightit Tufts University, Medford, Massachusetts 02155} \\
\r {38} {\eightit Waseda University, Tokyo 169, Japan} \\
\r {39} {\eightit University of Wisconsin, Madison, Wisconsin 53706} \\
\r {40} {\eightit Yale University, New Haven, Connecticut 06520} \\
\end{center}

\begin{abstract}
We present the results of a search for first generation scalar leptoquarks
(${\cal S}_1$) using
\hbox{110 $\pm$7 pb$^{-1}$} of  data  collected by the CDF experiment
 at Fermilab.  We search for $\lqlqbar$ pairs where both 
leptoquarks decay to an electron and a quark.
Three candidate events, 
with masses below 140 GeV/c$^2$ and consistent
with background expectations,  are observed.  
We obtain a \hbox{95\% C.L.} upper limit on the
production cross section as a function of the leptoquark mass.  Using a NLO
calculation of  $\lqlqbar$ production, we exclude scalar leptoquarks with 
mass less than 213 GeV$/c^2$  at 95\% C.L. 
for a branching ratio into $eq$  equal to 1.\\

\end{abstract}                           

\end{center}

\renewcommand{\baselinestretch}{2}

Leptoquarks are hypothetical color-triplet bosons, which carry both lepton and
baryon number and  appear in several extensions of the Standard Model
(SM)\cite{salam}.  Leptoquarks light enough to be produced at
current accelerators are
usually assumed to couple to quarks and leptons within the same generation, in
order to avoid large flavor-changing neutral current 
processes~\cite{buchmuller}.
First generation scalar leptoquarks (${\cal S}_1$) are 
assumed  to decay to an electron or positron and a first generation quark
 or antiquark with  branching ratio $\beta$.
Lower limits on
the leptoquark mass at 95\% confidence level (C.L.) have been reported 
by the CDF collaboration ($\mlq >$113 (80)~GeV/c$^2$ for
$\beta ({\cal S}_1\rightarrow eq$)=1 (0.5) from a data sample 
corresponding 
to 4.05 pb$^{-1}$ of integrated luminosity~\cite{cdf}), and by the D$\emptyset$
collaboration 
($\mlq >$ 133 (120)~GeV/c$^2$ for 
$\beta ({\cal S}_1 \rightarrow eq$)=1 (0.5) for an integrated luminosity
of 15 pb$^{-1}$~\cite{d0}).
Searches at \hbox{LEP-1} have excluded leptoquarks with masses below  
45~GeV/c$^2$ independent of the branching ratio~\cite{lep}. 
Limits from $ep$ colliders 
exclude masses up to 230 GeV/c$^2$~\cite{hera} for values of the
leptoquark-lepton-quark coupling $\lambda$ larger than 
$\sqrt {4 \pi \alpha_{em}}$.
The limits are weaker for smaller values of the coupling.
First generation leptoquarks with a mass of about 200 GeV/c$^2$ have been
suggested as a possible explanation for the recent observations
 of an excess of
events at very large values of the negative square of the 
momentum transfer Q$^2$ over the SM expectations by the  H1\cite{h1} and
 ZEUS\cite{zeus} experiments  at HERA. Such a hypothesis implies a 
100\% branching ratio into the electron-jet channel\cite{altarelli}.

Leptoquarks can be produced in pairs in $\ppbar$ collisions via  the strong
interaction through gluon-gluon fusion and $q\bar q$ annihilation. 
Production mediated by the leptoquark-lepton-quark coupling is
negligible, due to existing constraints on $\lambda$~\cite{hera} in the $\mlq$
region to which we are sensitive. 
The production cross section for a pair of scalar leptoquarks
 can therefore be calculated entirely within QCD,
and is known up to next-to-leading order (NLO) accuracy~\cite{kraemer}.

In this paper we present a search for first generation scalar leptoquarks
based on 110$\pm$7 pb$^{-1}$ of
 data collected with the Collider Detector at Fermilab (CDF) in $\ppbar$
collisions at $\roots$ = 1.8 TeV during the 1992-95 Tevatron runs.
 The distinctive signature for these events is two high-energy 
jets plus two isolated, high-energy electrons, where the invariant mass of  
each electron-jet system from a leptoquark decay corresponds to  the 
leptoquark mass.

   The CDF detector is described in detail 
in Ref.~\cite{detector}.
   The data sample for this analysis was collected 
with a high transverse-energy ($\Et$)
 central electron trigger ($\Et>$18 GeV, pseudorapidity
 $|\eta |<$ 1~\cite{eta}).
Electrons are detected with the central and plug electromagnetic and
hadronic calorimeters, the
central tracking chambers (CTC), and the central strip chambers (CES). 
Details of the electron selection can be found in Ref.~\cite{sacha}; the
requirements most relevant for this analysis are listed here.
We require two isolated electrons with $\Et >$ 25 GeV, 
one in the central calorimeter passing tight selection criteria,
 and a second one in the central or plug (1.1 $<|\eta |<$ 2.4) 
calorimeters and passing looser criteria. 
With respect to Ref.~\cite{sacha} we use a looser requirement of
 $E/P <$ 4 for both electrons, and
 a less stringent fiducial requirement on the second electron. 
The electron identification  efficiencies 
are measured from a sample of $Z \rightarrow e^+e^-$ to be 
86\%$\pm$3\%, 94\%$\pm$3\%, and 89\%$\pm$3\%
 for the tight central, loose central and 
loose plug selection, respectively. 

   Jet reconstruction is performed using a cone algorithm with radius $R$=0.7 
in ($\eta,\phi$) space. 
Corrections are applied for
energy lost in calorimeter cracks, energy outside the cone, and 
energy deposited inside the cone from the underlying event.
We require one jet with \hbox{$\Et >$ 30 GeV} and a second jet with
\hbox{$\Et >$ 15 GeV} in the pseudorapidity region  $|\eta |<$4.2.

   The main background in this search is due to the Drell-Yan process
 $Z/\gamma\rightarrow e^+e^-$
 plus two or more jets from initial state radiation.
Consequently, the invariant mass of the two electrons 
is required to be outside the $Z$ boson
mass window, \hbox{76 $<$ $M_{ee}$ $<$ 106 GeV/c$^2$}.
The radiated jets in Drell-Yan events are expected 
to be softer than the jets from ${\cal S}_1$ decays, in which the 
energies of the electrons and jets are similar. 
Therefore we require
\hbox{$\Sigma E_{T_{j1,j2}} = E_{T_{j1}}+E_{T_{j2}} > $ 70 GeV} and 
\hbox{$\Sigma E_{T_{e1,e2}} = E_{T_{e1}}+E_{T_{e2}} > $ 70 GeV} (denoted as the
$\Sigma \Et$ requirement).

\begin{figure}
\epsfysize=5in
\epsffile[36 162 540 684] {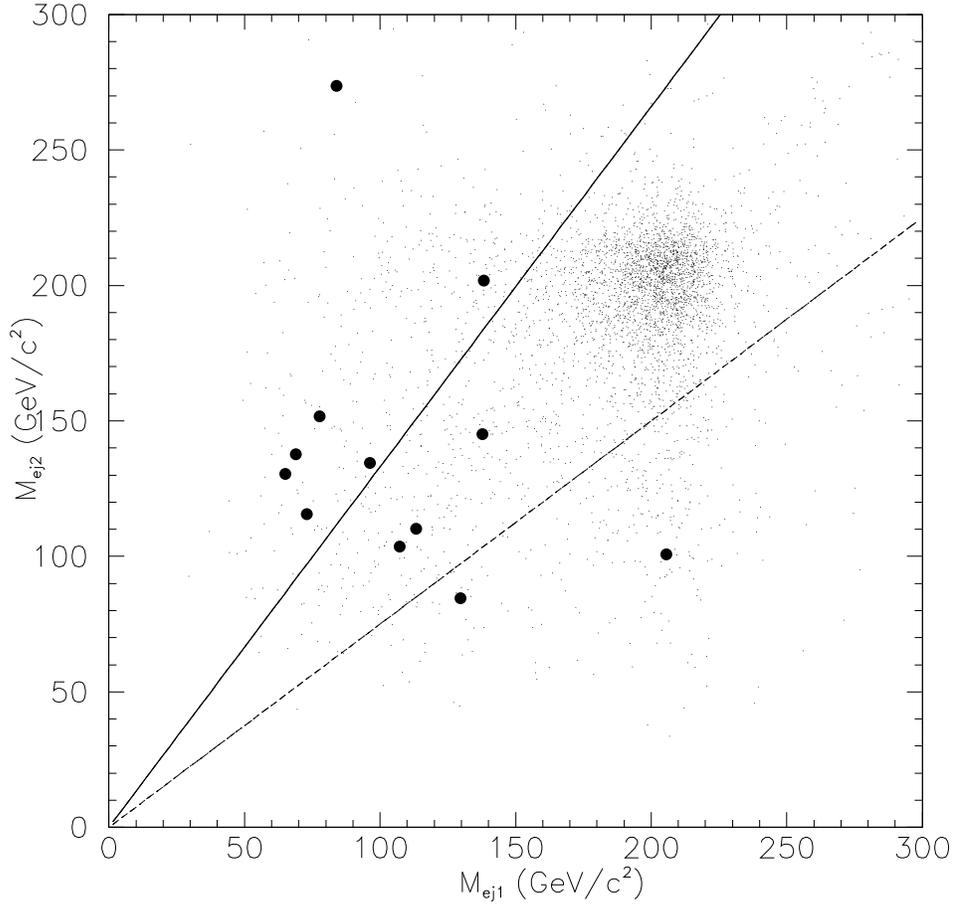}
\caption{Distribution of $M_{ej1}$ vs $M_{ej2}$ for events passing 
all requirements except the mass difference requirement, 
for data (full circles) and leptoquark Monte Carlo with 
$\mlq$ = 200 GeV/c$^2$ (dots).  The Monte Carlo events correspond to a total
integrated luminosity of 50 fb$^{-1}$.
  The lines define the region where the masses of the two
electron-jet pairs in the event are considered consistent with 
$\lqlqbar$ pair production.} 
\label{m12}
\end{figure}

\begin{figure}
\epsfysize=5in
\epsffile[36 162 540 684] {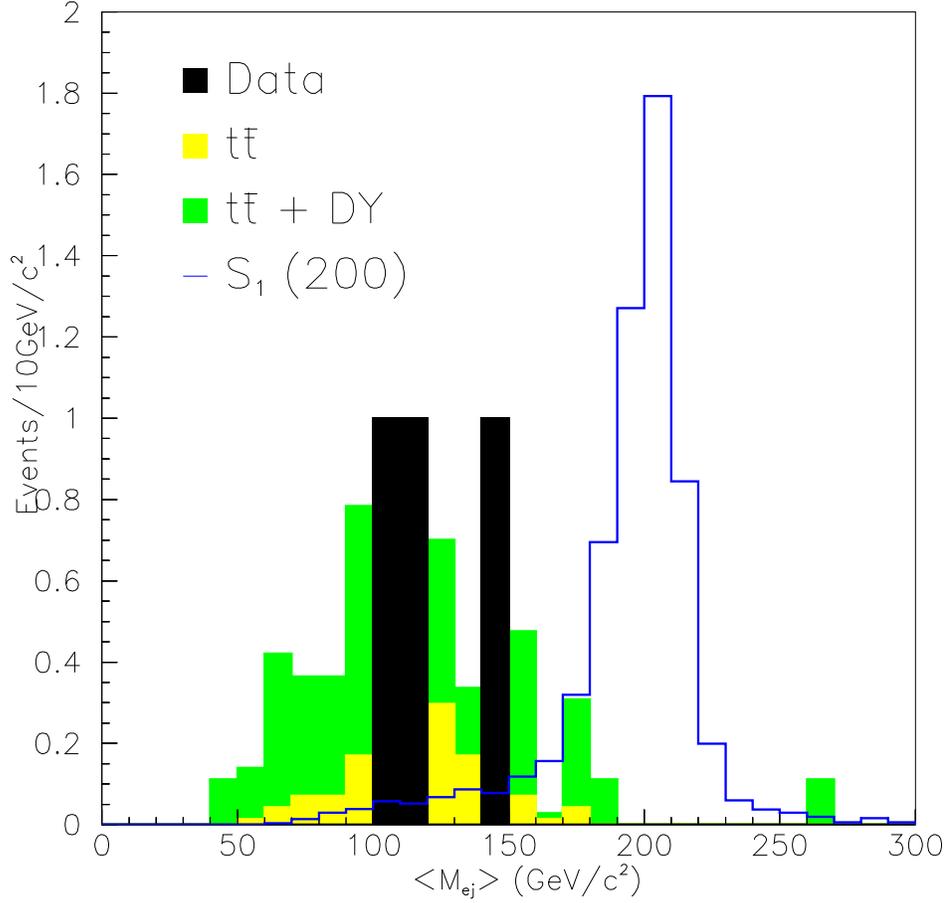}
\caption{Distribution of $\langle \mej \rangle $
for the data.
Superimposed are the  distributions for the $\ttbar$ and 
$\ttbar$ plus Drell-Yan backgrounds, normalized to the relative 
estimated numbers. The line histogram represents 
the expected $\langle \mej \rangle $ distribution for a 200 GeV/c$^2$ leptoquark
 for an integrated luminosity of 110 pb$^{-1}$, using the NLO cross section.}
\label{lqmean}
\end{figure}

   For the signal, the masses ($\mej$) of the two electron-jet pairs  
in each event are expected to be the same  within experimental resolution
in the absence of final state radiation.
   The invariant masses of the $e-$jet systems are reconstructed using the
two highest  $\Et$ electrons that pass our selection criteria and the two
highest  $\Et$ jets.
We resolve the  ambiguity in the electron-jet assignment by
choosing the pairing that gives the smallest mass 
difference between the two pairs.
Figure \ref{m12}  shows the scatter plot of 
the two electron-jet masses in each event passing the selection criteria
 for the data and for the Monte Carlo simulation of 200 GeV/c$^2$ leptoquark 
pairs (described below). 
Mis-assignment in the electron-jet pairing, due mainly 
to the presence of extra jets from
gluon radiation, results in events outside the mass peak.
To ensure that the candidate events  are consistent
with leptoquark production, we accept events with
 $M_{{ej}_1}$ and $M_{{ej}_2}$ in the region  defined by the 
two lines in Fig.~\ref{m12}, corresponding to two sigma
in  the mass difference resolution with respect to the
diagonal ($M_{{ej}_1}$ = $M_{{ej}_2}$).  
Three events pass the selection criteria.
Table 1 shows the number of events in the data
remaining after each cut. 
\begin{table}
\begin{center}
\begin{tabular}{lr}  
&Events\\
Inclusive Sample& 609K\\ \hline
Cut&\\ \hline
$E_{T_{e1,e2}} >$ 25 GeV&7466  \\
$E_{T_{j1,j2}}>$ 30, 15 GeV &228\\
$M_{ee}$ cut& 27\\
$\Sigma E_{T_{e1,e2}}, \ \Sigma E_{T_{j1,j2}} > $70 GeV &12\\
$M_{e,jet}$ cuts&3\\
\end{tabular}
\end{center}
\label{table:data}
\caption{Number of events  in the data passing the kinematical cuts.
} 
\end{table}
   For events surviving all selection requirements, we calculate 
the mean of the two electron-jet
pair masses, $\langle \mej \rangle$. The largest mean mass of the 
three candidates is 140 GeV/c$^2$.
Figure \ref{lqmean} shows the $\langle \mej   \rangle$ distribution 
for the three events in the data, 
together with predicted  distributions 
for Drell-Yan and $\ttbar$ backgrounds (described below), 
normalized to the respective
number of expected events. Also shown is the expected distribution
from a Monte Carlo simulation of 200 GeV/c$^2$ leptoquark production, 
normalized to 
the integrated luminosity using
the theoretical cross section~\cite{kraemer}.
Of the three observed events one has two photons that are interpreted as 
neutral jets, therefore satisfying the selection criteria,  
and  one has a jet tagged as a $b$ quark, which is consistent with
this event originating from $\ttbar$ background.
 
   The kinematical acceptance for leptoquark masses in the range
140 to 240 GeV/c$^2$  was studied
with  samples of 10000 leptoquark pairs
decaying into $eq$ produced with the PYTHIA generator\cite{pythia} and 
passed through a detailed detector simulation.
The samples
were generated using 
the CTEQ4L \cite{cteq} parton distribution functions, with the renormalization
and factorization scale \hbox{ Q$^2$ = $\mlq ^2$}.
The acceptance after the electron  $\Et$ and isolation requirements
 varies from 56\% to 65\% for 
leptoquark masses of 140 to  240 GeV/c$^2$. 
The jet  $\Et$ requirement reduces the leptoquark acceptance to 
49\% to 58\% for the same $\mlq$ range.
After the $M_{ee}$ and $\Sigma \Et$ requirements the acceptance varies 
from 40\% to
54\%, and after the mass difference requirement from 33\% to 44\%.
The acceptance estimates were checked by releasing the requirements on the jet
transverse energies and comparing the value of the production cross section
of the $Z$ boson, as a function of the number of jets in the event,
 with the published values~\cite{sigmaz}.

   The Drell-Yan background  
is estimated using the PYTHIA Monte Carlo program.
The Drell-Yan sample is normalized  to the number of events in the data 
in the  76 $< M_{ee} <$ 106 GeV/c$^2$ interval 
with  at least one jet with $\Et >$ 30 GeV and a second one
with $\Et >$ 15 GeV.
The expected background is estimated to be 12.1$\pm$2.9 events before the 
mass difference requirement and 
4.4$\pm$2.2 events after this  requirement. These estimates are in agreement
with the results from an exact matrix element
calculation for the  Drell-Yan plus two jets cross section.
   The amount of background from $\ttbar \rightarrow W^+ W^- \bbar$ production
when both $W$ bosons  decay to $e\nu$ is estimated to be 
2.2$\pm$0.5 events before 
and 1.4$\pm$0.3 events after the  mass difference requirement. 
The $\ttbar$ sample was generated with PYTHIA, and normalized  with the
CDF measured $\ttbar$ cross section of 7.5$^{+1.9}_{-1.6}$ pb~\cite{top}.

Other backgrounds, from ${b\bar{b}}$ and  Z$\rightarrow \tau^+ \tau^- $, are
negligible due to the electron isolation  and large
transverse energy requirements on the electron and jets. 

   We have considered systematic uncertainties  on the 
acceptance from the following sources: gluon radiation (9\%), 
choice of the parton distribution functions (5\%), 
electron identification efficiency (4\%),
jet energy scale (2\%), and Monte Carlo statistics (2\%).
 The  resulting total systematic uncertainty  is 13\%,
including the 7\% uncertainty on the luminosity. 

\begin{figure}
\epsfysize=5in
\epsffile[36 162 540 684] {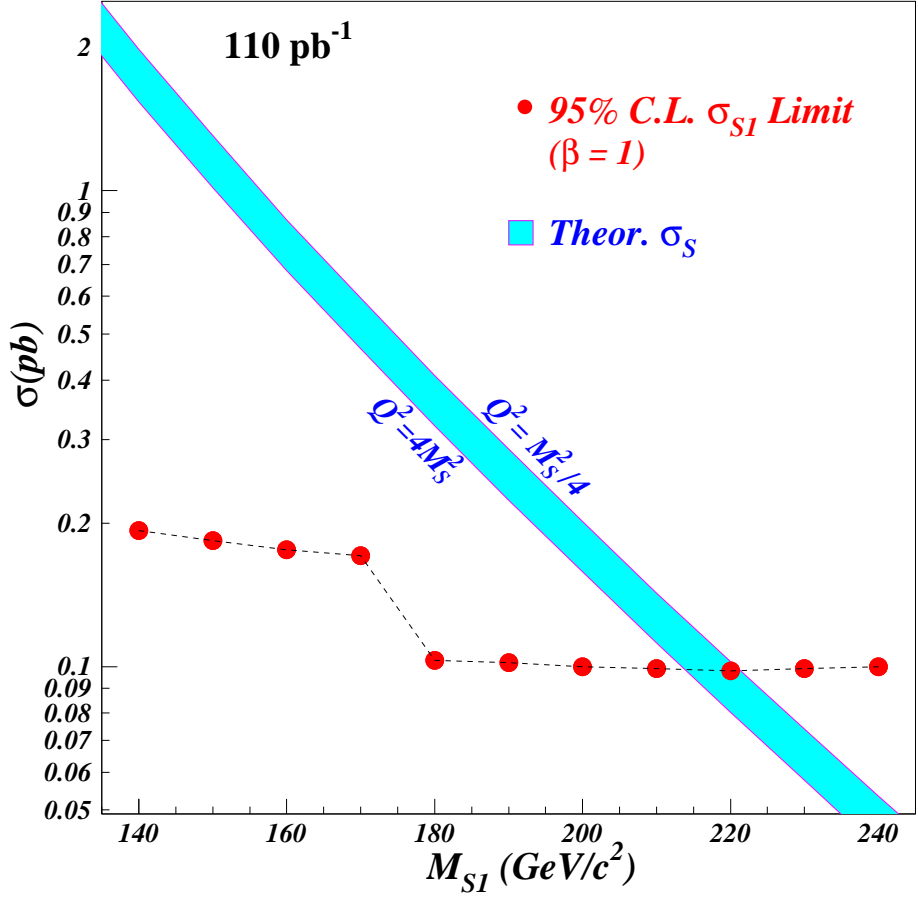}
\caption{ The  95\% C.L. upper limit on the cross section 
for $\lq$ pair production, assuming $\beta$=1.
The band represents the spread of the NLO calculation of 
$\sigma _{\lqlqbar }$
of Ref.~\protect\cite{kraemer} 
 as a function of the Q$^2$ scale, with
the CTEQ4M parton distribution functions.} 
\label{sl_mean_pt}
\end{figure}

   The data shown in Fig.~\ref{lqmean} are  used to set a limit on 
the $\lqlqbar$ production cross section versus $\mlq$. 
The number of candidates for a given leptoquark mass is defined as the number 
of observed events with $\langle \mej \rangle$ in a $\pm 3\sigma$  interval
around that mass, $\sigma$ being the 
experimental resolution. 
We take all observed  events to be candidates to obtain a conservative 
95\% C.L. upper limit on the cross section as a function of $\mlq$.
The limit calculation accounts for the statistical uncertainty on the
number of observed events and the total systematic uncertainty.
The values of the 95\% C.L. limits for $\sigma (\lqlqbar )$ are 
listed in Table 2 for
different leptoquark masses, together with the  final acceptance times 
electron identification efficiency. Also listed are the 
next-to-leading order theoretical
calculation of the cross sections for pair production of scalar
leptoquarks at the Tevatron, calculated with the CTEQ4M 
parton distribution functions for two choices of the 
Q$^2$ scale~\cite{kraemer}.
For a leptoquark mass of 200 GeV/c$^2$ we obtain a limit on the
cross section of 0.1 pb.
The same limits are shown in Fig.~\ref{sl_mean_pt}, for 
$\beta({\cal S}_1 \rightarrow eq)$ = 1, along 
with the theoretical cross section expectations.
By using these estimates of $\sigma (\lqlqbar )$, we obtain a lower limit for
$\mlq$ of 213 GeV/c$^2$ for $\beta$ =1 at the 95\% C.L.
\begin{table}
\begin{center}
\begin{tabular}{c|c|c|cc} 
$\mlq$(GeV/c$^2)$ & 95\% CL $\sigma (\lqlqbar )$(pb) &Acc.$\times$ eff.&
 \multicolumn{2}{c}{$\sigma(\lqlqbar )_{theor.}$(pb)  } \\ 
 &($\beta$=1) && Q$^2$=$\mlq^2/4$ & 
Q$^2$=4$\mlq ^2$ \\
\hline 
140 & 0.19 & 0.23 &1.98 & 1.54 \\
150 & 0.18 & 0.24 &1.30 & 1.01 \\
160 & 0.18 & 0.25 &0.87 & 0.68 \\
170 & 0.17 & 0.26 &0.59 & 0.46 \\
180 & 0.10 & 0.27 &0.41 & 0.32 \\
190 & 0.10 & 0.27 &0.29 & 0.22 \\
200 & 0.10 & 0.28 &0.20 & 0.16 \\
210 & 0.10 & 0.28 &0.14 & 0.11 \\
220 & 0.10 & 0.28 &0.10 & 0.08 \\
230 & 0.10 & 0.28 &0.07 & 0.06 \\
240 & 0.10 & 0.28 &0.05 & 0.04 \\
\end{tabular}
\end{center}
\caption{The experimental $\lqlqbar$ cross section limit, acceptance $\times$ 
electron identification efficiency, and the theoretical cross section
versus $\mlq$.}
\label{table:xsecl}
\end{table}

   In conclusion, we have presented a search for first 
generation ${\cal S}_1$ pair production with the CDF experiment.
Three events, consistent with SM background 
expectations, have been observed. 
Limits on the $\lqlqbar$ production cross section as a 
function of mass are obtained.
The existence of scalar
leptoquarks with mass below 213 GeV/c$^2$ for 
$\beta$ =1 is excluded  at the 95\% C.L.

     We thank the Fermilab staff and the technical staffs of the
participating institutions for their contributions.  This work was
supported by the U.S. Department of Energy and National Science Foundation;
the Italian Istituto Nazionale di Fisica Nucleare; the Ministry of Science,
Culture, and Education of Japan; the Natural Sciences and Engineering Research
Council of Canada; the National Science Council of the Republic of China;
 and the A. P. Sloan Foundation.


\begin{thebibliography}{99}

\bibitem{salam} J.~C.~Pati and A.~Salam, Phys. Rev. {\bf D19}, 275 (1974);
 H.~Georgi and S.~Glashow, Phys. Rev. Lett. {\bf 32}, 438 (1974); 
 L.~F.~Abott and E.~Farhi, Nucl. Phys. {\bf B189}, 547 (1981); 
 E.~Eichten {\em et al.}, Phys. Rev. Lett. {\bf 50}, 811 (1983); 
 E.~Witten, Nucl. Phys. {\bf 258}, 75 (1985); 
 M.~Dine {\em et al.}, Nucl. Phys. {\bf B259}, 519 (1985); 
 J.~Breit {\em et al.}, Phys. Lett. {\bf 158B}, 33 (1985);
 S.~Pakvasa, Int. J. Mod. Phys. {\bf A2}, 1317 (1987); 
 J.~L.~Hewett and T.~G.~Rizzo, Phys. Rep. {\bf 183}, 193 (1989); 
 J.~L.~Hewett and S.~Pakvasa, Phys. Rev. {\bf D37}, 3165 (1988);
 P.~H. Frampton, hep-ph/9706220.

\bibitem{buchmuller} W.Buchm\"uller and D. Wyler, Phys. Lett. {\bf B177},
337 (1986).

\bibitem{cdf} F.~Abe {\em et al.}, The CDF Collaboration, 
Phys. Rev. D, {\bf 48}, R3939 (1993).

\bibitem{d0} S.~Abachi {\em et al.}, The D$\emptyset$ Collaboration, 
 Phys. Rev. Lett. {\bf 72}, 965 (1994).

\bibitem{lep} B.~Adeva {\em et al.}, Phys. Lett. {\bf B261}, 169 (1991); 
G.~Alexander {\em et al.}, Phys. Lett. {\bf B263}, 123 (1991);
P.~Abreu {\em et al.}, Phys. Lett. {\bf B275}, 222 (1992);
D.~Decamp {\em et al.}, Phys. Rep. {\bf C216}, 253 (1992).


\bibitem{hera} 
M. Abbiendi {\em et al.}, The ZEUS Collaboration, Phys. Lett. {\bf B306},
 173 (1993);
S. Aid {\em et al.}, The H1 Collaboration, Phys. Lett. {\bf B369}, 173 (1996).

\bibitem{altarelli} G.~Altarelli {\em et al.}, hep-ph/9703276; 
J.~Bl\"umlein, hep-ph/9703287; 
K.~S.~Babu {\em et al.}, hep-ph/9703299; 
J.~L.~Hewett and T.~Rizzo, hep-ph/9703337.

\bibitem{h1} C.~Adloff {\em et al.}, The H1 Collaboration, 
Z.~Phys. {\bf C74}, 191 (1997).  

\bibitem{zeus} J.~Breitweg {\em et al.}, The 
ZEUS Collaboration, Z.~Phys. {\bf C74}, 207 (1997). 

\bibitem{kraemer} M. Kr\"{a}mer {\em et al.}, 
Phys. Rev. Lett. {\bf 79}, 341(1997).

\bibitem{detector} F.~Abe {\em  et al.},  Nucl. Instrum. and
Methods Phys. Res. {\bf A271}, 387 (1988).

\bibitem{eta} The pseudorapidity $\eta$  is defined as $-\ln \tan(\theta/2)$,
 where $\theta$  is the polar angle with respect to the proton direction;
$\phi$ is the azimuthal angle.  The transverse momentum of a particle 
is defined as $\Pt = P \sin \theta$, where P is the momentum of the track
associated with the particle as measured in the CTC. Similarly, 
$\Et$ is defined as 
$\Et = E \sin \theta $, where $E$ is the energy associated with the particle 
as measured in the calorimeters. 

\bibitem{sacha} F.~Abe {\em et al.}, The CDF Collaboration, 
Phys. Rev. {\bf D52}, 2624 (1995).


\bibitem{pythia} T.~Sj\"{o}strand, Comput. Phys. Commun. {\bf 82}, 74 (1994).
We use version 6.1. 

\bibitem{cteq} H. L. Lai {\em et al.}, The CTEQ collaboration, 
Phys. Rev. {\bf D55}, 1280 (1997).

\bibitem{sigmaz} F. Abe {\em et al.}, The CDF Collaboration, 
Phys. Rev. Lett. {\bf 76}, 3070 (1996).

\bibitem{top} P. Tipton, Proceedings of the
XXVIII International Conference on High Energy Physics
   Warsaw, Poland, 1996.
 

\end{thebibliography}
\end{document}